# Graphene-based qubits in quantum communications

G. Y. Wu[1,2,*] and N.-Y. Lue[1]

[1]Department of Physics, National Tsing-Hua University, Hsin-Chu 30013, Taiwan, ROC; [2]Department of Electrical Engineering, National Tsing-Hua University, Hsin-Chu 30013, Taiwan, ROC; [*]E-mail: yswu@ee.nthu.edu.tw.

**Abstract**

We explore the potential application of graphene-based qubits in photonic quantum communications. In particular, the valley pair qubit in double quantum dots of gapped graphene is investigated as a quantum memory in the implementation of quantum repeaters. For the application envisioned here, our work extends the recent study of the qubit (Wu *et al.*, arXiv: 1104.0443; Phys. Rev. B **84**, 195463 (2011)) to the case where the qubit is placed in an in-plane magnetic field configuration. It develops, for the configuration, a method of qubit manipulation, based on a unique AC electric field-induced, valley-orbit interaction-derived mechanism in gapped graphene. It also studies the optical response of graphene quantum dots in the configuration, in terms of valley excitation with respect to photonic polarization, and illustrates faithful photon ↔ valley quantum state transfers. This work suggests the interesting prospect of an all-graphene approach for the solid state components of a quantum network, e.g., quantum computers and quantum memories in communications.

# I. Introduction

Quantum bits (qubits) are the fundamental units of quantum information exchanged in quantum communications (QCs) [1,2] or processed in quantum computing [3]. Apart from the flying photon qubit which plays an essential role in QCs, of particular interest among the qubits proposed are the static, solid state ones that utilize the spin [4] or valley [5] degrees of freedom of electrons. Such qubits can be used for storage of quantum information and, moreover, having the structure of gated devices, may be scalable and electrically manipulated, similar to semiconductor IC transistors.

The present work focuses on the potential application of valley-based qubits in QCs, which is based on the unique physical properties of graphene recently discovered [6] and extensively studied [7]. As is well known, graphene is a two-dimensional material of hexagonal lattice, with a distinctive band structure characteristic of a Dirac particle. More importantly, there are two independent energy valleys located, respectively, at K and K' of the Brillouin zone. A low-lying charge carrier may sit in either of the valleys and is endowed with a binary-valued degree of freedom (d.o.f.) analogous to spin.

It has been conjectured for some time that this valley d.o.f. is suited to the coding of quantum information,[8] and the conjecture is recently realized in the proposal of Reference 5 by Wu *et al*. It is shown that a valley-based qubit (called valley pair qubit) can be implemented by utilizing two coupled quantum dots (QDs) in gapped graphene (epitaxially grown on SiC [9] or BN [10], for example). As explained in Reference 5, a valley pair qubit is basically a two-electron system in the double quantum dots (DQD), with the state space consisting of "valley singlet/triplet states" representing, respectively, logical 0/1 values.

Reference 5 develops, for the valley pair qubit, a method of quantum state manipulation suited to the implementation of valley-based quantum computing. It employs a static tilted magnetic field configuration, where the in-plane field freezes the electron spin while the normal field induces an asymmetry between K and K' valleys, creating a corresponding "valley Zeeman splitting" [11,12]. A key element of the method is that the splitting in each QD of the qubit can be tuned independently (with a gate voltage) to create across the qubit DQD a differentiation in the size of splitting, which drives a state transformation for the qubit manipulation. The physics underlying the electric tuning of valley splitting involves a unique, relativistic type physical mechanism in gapped graphene, namely, the following valley-orbit interaction (VOI) ($\tau_v$ = +1 / -1 for K / K', $2\Delta$ = band gap, $m^*$ = electron effective mass, $V$ = potential energy, $\bm{p}$ = momentum operator)

$$H_{VOI} = \tau_v \frac{\hbar}{4m^*\Delta} \nabla V \times \vec{p}.$$

The VOI is an analogue of the Rashba mechanism [13] of spin-orbit interaction (SOI). While the SOI has been demonstrated an effective mechanism for electrical manipulation of spin qubits in semiconductors,[14] the VOI provides an alternative mechanism in the case of graphene where the SOI strength is known to be weak [7].

The present work belongs to the series of our recent theoretical investigations of valley pair qubits, and extends the scope of potential applications for valley pair qubits from quantum computing to photon-based QCs. In particular, it examines the issue of photon ↔ valley quantum state transfer (QST) critical to the application envisioned here, and discusses the feasibility of valley-based quantum memories in the implementation of quantum repeaters for photonic QCs.

It is well known that, with the racing speed of light, photonic QCs hold great promise for quantum networks or long distance distributions of quantum keys in quantum cryptography [15]. In these applications, it is essential to generate with photonic signals a long range quantum entanglement between two sites. However, due to the exponential decay of photonic signals in the channel, the entanglement usually attenuates with distance, making the long range distribution of entanglement a challenging task. The quantum repeater protocol is a strategy that solves the problem of attenuation by dividing the channel into many segments and distributing the entanglement in a cascading fashion.[2,16] With the protocol, quantum entanglement is generated in each segment and then connected with that in the adjacent segment (by entanglement swapping [17]). The same process is applied over and again, each time with the entanglement range being doubled, until eventually it is expanded far enough to cover the two parties (sender and receiver) in the communication. In the protocol, photons are utilized to carry quantum entanglement (over a distance less than the light attenuation length), and solid state qubits are utilized as quantum memories to temporarily store the entanglements already established in the segments. Since the entanglements are built in a probabilistic manner, their storage in solid state qubits is of vital importance in that it synchronizes the entanglements for swapping.

One of the advantages in using solid state quantum memories, such as a semiconductor-[18] or the graphene- based one envisioned here, lies in the accessibility of quantum state manipulation via electrical gate control in the above cascading process. However, important issues arise. For example, an elementary and frequent operation in quantum repeaters with graphene-based quantum memories would be the conversion of a quantum state, from a photonic form to a valley-based one in graphene and vice versa, and it is crucial to minimize the quantum distortion resulting from such quantum state transfers (QSTs). This places a constraint on the working configuration of valley-based qubits, as well as on the corresponding method of state manipulation. The work presented below addresses these important issues.

In the work, we focus primarily on photonic QCs using the photonic polarization ($\sigma+$ / $\sigma-$) for coding. The constraint imposed by a faithful QST then requires that the optical response of graphene QDs be symmetric, in terms of valley excitation with respect to the photonic polarization. In order to see how the constraint arises, consider the following simple example where the valley-based qubit comprises only a single QD-confined electron, with the quantum information being encoded in the linear combination of $|K>$ and $|K'>$ (of the electron state). Although the feasibility of quantum information processing based on such a simple qubit has not yet been demonstrated, it provides nonetheless a simple illustration. In this case, a faithful QST from photon to valley qubits means

$\alpha|\sigma+> + \beta|\sigma-> \rightarrow \alpha|K> + \beta|K'>$,

which can occur, if the QD reacts to an incoming photon symmetrically as follows,

$|K(valence)> + |\sigma+> \rightarrow |K(conduction)>$   (with amplitude M),           (I-1)

$|K'(valence)> + |\sigma-> \rightarrow |K'(conduction)>$   (with amplitude M'),

$|M| = |M'|$.

In fact, in the absence of a normal magnetic field, the optical excitation in gapped graphene is indeed symmetric, and obeys the selection rule in (I-1) approximately, as pointed out previously. [19] With (I-1), it follows that

α|σ+> + β|σ-> → α $e^{i\chi}$ |K(conduction)> + β |K'(conduction)>     ($e^{i\chi}$ = M / M'),

which is obtained by superposing the two quantum processes in (I-1). A further valley state manipulation can be applied to annihilate the extra phase χ appearing in the state, thus achieving a faithful QST from photon to valley qubits. The criterion of a symmetric optical response demonstrated above applies as well to the valley pair qubit considered in the work, as shall soon become clear when we discuss the QST between photon and valley pair qubits.

While the tilted magnetic field configuration works perfectly for quantum computing [5], the presence of the normal magnetic field component breaks valley symmetry and forbids a symmetric optical response. A focus of the present work is to investigate valley pair qubits in the alternative, in-plane magnetic field configuration, and develop a corresponding method of qubit manipulation in the case.

The presentation is organized as follows. In **Sec. II**, we provide a description of valley pair qubits in the in-plane magnetic field configuration. In **Sec. III**, we present the Schrodinger type equation for electron states in gapped graphene. The equation incorporates relativistic type corrections (R.C.), including the VOI, up to the $2^{nd}$ order, and is employed to study an AC electric field-induced, VOI-based mechanism. Following the mechanism, the method of state manipulation is developed for valley pair qubits in the in-plane magnetic field configuration. It is shown that the VOI-based mechanism is a $2^{nd}$-order relativistic type effect, and an estimate of the effect gives the time scale of O(10ns) for the qubit manipulation. In **Sec. IV**, we discuss the optical response of graphene QDs, and illustrate the faithful QST from photon to valley pair qubits and vice versa. We also consider the back-to-back QST such as valley → photon → valley, and show that the QST is highly faithful. In **Sec. V**, we summarize our findings. In **Appendix A**, we present the derivation of the Schrodinger type equation in gapped graphene, with the relativistic type corrections included up to the $2^{nd}$ order. In **Appendix B**, we provide the mathematical details involved in the derivation of the AC electric field-induced, VOI-based effect. Finally, in **Appendix C**, we estimate the coherence time of valley pair qubits in the in-plane magnetic field configuration, and show that it can be made long enough for the VOI-based qubit manipulation developed in **Sec. III**.

**II. Valley pair qubits**

The pair of coupled QDs (in the x-y plane) for the qubit may be formed by spatially modulating graphene energy bands, e.g., via back gate voltages, to provide a band gap-caused quantum confinement. As shown in Fig. 1a, electrical gates, $V_L$, $V_R$, and $V_c$, are also placed near the QDs to further modulate the QD confinement potential. In addition, $V_c$ controls the potential barrier between the two QDs and, hence, the corresponding tunneling amplitude $t_{d-d}$, too. The state of a confined electron is characterized by the following set of indices, $(n, X, \tau_v, s_x)$. Here, n = QD energy level index, X = L or R, denoting the left / right QD, and $s_x = \pm\frac{1}{2}$ being the electron spin component in the x-direction. A static in-plane magnetic field is applied, which freezes the spin degree of freedom at, for example, $s_x = \frac{1}{2}$, as shown in Fig. 1b.

The Fermi energy is set at such a level that a population of two electrons resides in the DQD structure, interacting with each other with the on-site Coulomb repulsion energy (U). The valley pair qubit operates in the low-energy charge configuration ($1_L$, $1_R$), where the two electrons are separately confined in the QDs with the following exchange-type effective interaction between the electrons,

$$H_J = \frac{1}{4} J \vec{\tau}_L \cdot \vec{\tau}_R,$$

with J ~ 4 $t_{d-d}^2$/(U- δε) being the exchange integral, and $\tau_{L(R)}$ = "Pauli valley operator" (identical to Pauli spin operator). δε here refers to the energy detuning between (n = 0, X = L, $\tau_v$, $s_x$ = ½) and (n = 0, X = R, $\tau_v$, $s_x$ = ½). Note that J is electrically tunable, through the adjustment of $t_{d-d}$ via $V_c$, or that of δε via back gate voltages. The eigenstates of $H_J$ are the following valley singlet/triplet states:

$$|z_S> = \frac{1}{\sqrt{2}} (c_{K_L}^+ c_{K_R'}^+ - c_{K_L'}^+ c_{K_R}^+) | vaccum >,$$
$$|z_{T_0}> = \frac{1}{\sqrt{2}} (c_{K_L}^+ c_{K_R'}^+ + c_{K_L'}^+ c_{K_R}^+) | vaccum >,$$
$$|z_{T_+}> = c_{K_L}^+ c_{K_R}^+ | vaccum >,$$
$$|z_{T_-}> = c_{K_L'}^+ c_{K_R'}^+ | vaccum >.$$

An exchange splitting (= J) exists between the singlet $|z_S>$ and the triplet {$|z_{T0}>$, $|z_{T+}>$, $|z_{T-}>$}, as shown in Fig. 1c. In the above, $K_L$ = (n = 0, L, $\tau_v$ = 1), and $c_{KL}^+$ denotes the corresponding electron creation operator. Other notations are similarly defined.

The Hilbert space expanded by $|z_S>$ and $|z_{T0}>$ constitutes the qubit state space (denoted as $\Gamma_v$). $\Gamma_v$ is isomorphic to the spin-1/2 state space and, hence, can be represented by the surface of a sphere (i.e., the Bloch sphere). $|z_{T+}\rangle$ and $|z_{T-}\rangle$ are outside $\Gamma_v$ and not needed in the application of quantum computing/communications. Physically, they are coupled to $|z_S\rangle$ and $|z_{T0}\rangle$ by the intervalley scattering K ↔ K', and provide a channel of leakage contributing to qubit decoherence. In **Appendix C**, the intervalley scattering of a QD-confined electron is considered. [20]

Valley pair qubits can be manipulated all electrically. For example, if the exchange coupling J is maintained for a duration of $t_z$, it produces the unitary transformation, $R_z(\theta_z)$, i.e., a rotation about the z-axis of the Bloch sphere with the angle of rotation $\theta_z = J t_z / \hbar$. [5] However, in order to manipulate the qubit to an arbitrary point on the Bloch sphere, we need, in addition to $R_z$, a second independent state transformation. **Sec. III** addresses this important issue, and shows how to produce a rotation about the x-axis of the Bloch sphere (called $R_x$ below).

Quantum states of valley pair qubits are analogous to spin singlet/triplet states in the spin pair scheme [21,22]. As such, valley pair qubits are characterized by the same distinctive advantages provided in the scheme, e.g., scalability and decoherence-free state space. The method developed in the scheme for initialization / readout / two-bit qugate (CPHASE) operation, as described by Taylor et al., [14] may also be adapted here. With the method and the single qubit operations $R_x$ and $R_z$, universal quantum computing [23] can be achieved using valley-pair qubits.

**III. VOI-based state manipulation**

We note that the valley pair states, $|x_-\rangle$ and $|x_+\rangle$, defined below,

$$|x_-\rangle = \frac{1}{\sqrt{2}}(|z_{T_0}\rangle - |z_S\rangle) = c_{K_L'}^+ c_{K_R}^+ |vaccum\rangle,$$

$$|x_+\rangle = \frac{1}{\sqrt{2}}(|z_{T_0}\rangle + |z_S\rangle) = c_{K_L}^+ c_{K_R'}^+ |vaccum\rangle,$$

correspond to $|s_x = -1/2\rangle$ and $|s_x = 1/2\rangle$, respectively, as implied by the isomorphism between $\Gamma_v$ and the spin-1/2 state space. Comparing the expressions of $|x_-\rangle$ and $|x_+\rangle$, one sees that if one can break the symmetry between K ($\tau_v = 1$) and K' ($\tau_v = -1$) in one or both of the QDs, then a differentiation may be created between $|x_-\rangle$ and $|x_+\rangle$ leading to the following contrasting time evolution, namely,

$$|x_-\rangle \rightarrow e^{-i\Phi}|x_-\rangle,$$
$$|x_+\rangle \rightarrow e^{i\Phi}|x_+\rangle,$$

i.e., a rotation about the x-axis of the Bloch sphere.

There are two approaches to produce the needed valley asymmetry. Firstly, a normal magnetic field may be applied, as in the proposal of Wu et al. for quantum computing [5]. For QCs, if the same approach is employed, the field would have to be switched on and off frequently (e.g., off during the photon ↔ valley QST, in order to achieve a faithful QST, and on when the valley qubit is being processed in the quantum repeater) at the same frequency used in sending/receiving the photonic signal, leading to complications in the application. Or, alternatively, one may resort to the second approach where an in-plane magnetic field configuration is employed. In the following, we consider a QD-confined electron in this configuration, and show that an AC electric field can replace the normal magnetic field and induce the required valley asymmetry. This is termed the AC electric field-induced, VOI-based effect, and it enables a rotation about the x-axis of the Bloch sphere.

**The QD profile**

Apart from the AC electric field, the VOI-based effect depends also on the QD confinement potential. We describe briefly this dependence here. Let $V_{QD}$ be the QD confinement potential. Two profiles are considered. In one case, $V_{QD} = V_2(x,y) + V_3(x)$, with $V_2(x,y) = \frac{1}{2} m^* w_0^2 (x^2 + s^2 y^2)$ and $V_3(x) = 1/3 \ m^* w_0^2 \ k_{3x} x^3$. "s" in $V_2$ parametrizes the anisotropy of a generic quadratic potential and is taken to be of the order of unity. "$k_{3x}$" in $V_3$ characterizes the strength of the cubic potential. In the other case, $V_{QD} = V_2(x,y) + e\ \varepsilon_x\ x + V_4(x)$, involving an electric field ($\varepsilon_x$) in the x-direction and a quartic potential $V_4(x) = 1/4 \ m^* w_0^2 \ k_{4x} x^4$. "$k_{4x}$" in $V_4$ characterizes the strength of the quartic potential, and $\varepsilon_x$ may be produced by gate $V_c$ in Fig. 1a. In either case, in the presence of an AC electric field, the total potential energy of the electron is taken to be $V(x,y) = V_{QD} + V_{ac}$, with $V_{ac} = e\ \varepsilon_y sin(w_s t)\ y$ being the time-dependent potential due to the AC electric field. Here, the corresponding AC electric field, $\varepsilon_y sin(w_s t)$, is taken in the y-direction, and may be produced by gate $V_L$ or $V_R$ in Fig. 1a.

As shall become clear below, in the cubic case, the AC electric field-induced, VOI-based effect scales with $k_{3x}$ (and thus vanishes in the absence of $V_3$). Therefore, the presence of $V_3$ in $V_{QD}$ is an important requirement here. In the quartic case, the electric field $\varepsilon_x$ displaces the electron to a new equilibrium position ($x_0 = e\ \varepsilon_x / m^* w_0^2$), and a cubic term appears when $V_{QD}$ is expanded around $x_0$. This also enables the VOI-based effect.

**The Schrodinger type equation with "relativistic correction" up to the 2$^{nd}$ order**

Generally the two-band model (i.e. the Dirac equation) is a good description of both the conduction and valence bands in graphene. [7] However, in order to facilitate an analytical study of the VOI-based effect, we focus here on the regime where the electron is near the conduction band edge, i.e., $E/\Delta \ll 1$ where E = electron energy with respect to the band edge. (The study can easily be extended to near-band-edge valence band holes.) In the regime, the Dirac equation is reduced to the Schrodinger type equation derived in **Appendix A**. In the cubic case where $V_{QD} = V_2(x,y) + V_3(x)$, we have

$$H(x,y,t)\psi(x,y,t) = i\hbar\partial_t \psi(x,y,t),$$

$$H(x,y,t) = H^{(0)}(x,y,t) + V_3(x) + H^{(1)}(x,y,t) + H^{(2)}(x,y,t), \quad \text{(III-1)}$$

$$H^{(0)}(x,y,t) = \frac{\vec{p}^2}{2m^*} + V_2(x, y + y_0(t)) - \frac{1}{2}m^* w_0^2 y_0^2(t), \quad y_0(t) = \frac{e\varepsilon_y \sin w_s t}{m^* w_0^2}.$$

$y_0(t)$ here is the time-dependent electron displacement due to the AC electric field.

Eqn. (III-1) is correct to $O(E/\Delta)^2$. $H^{(0)}$ in the Hamiltonian describes a standard quantum harmonic oscillator (QHO). We consider the weak field limit where $|y_0| \ll Y$ ($Y = (\hbar/m^* w_0)^{1/2}$ being the size of the QHO oscillation amplitude) and ignore the $y_0^2$ term in $H^{(0)}$. This linearization does not affect the discussion below, since as shall become clear, the leading order of VOI-based effect is linear in $y_0$. $H^{(0)} + V_3$ constitutes the "nonrelativistic" part of the Hamiltonian. $H^{(1)}$ and $H^{(2)}$ are, respectively, the 1st- and 2nd-order relativistic type corrections, with $\|H^{(1)}\| / \|H^{(0)} + V_3\| \sim O(E/\Delta)$ and $\|H^{(2)}\| / \|H^{(0)} + V_3\| \sim O(E^2/\Delta^2)$.

We separate, in $H^{(1)}$ and $H^{(2)}$, valley –dependent and –independent terms. Specifically, we write

$$H^{(1)} = H_0^{(1)} + H_\tau^{(1)}, \qquad (\text{III-1.1})$$

$$H_0^{(1)} = -\frac{\vec{p}^4}{8m^{*2}\Delta} - \frac{1}{8m^*\Delta}(\vec{p}^2 V_2) - \frac{1}{8m^*\Delta}(\vec{p}^2 V_3),$$

$$H_\tau^{(1)} = \tau_v \frac{\hbar}{4m^*\Delta}[\nabla V_2(x, y+y_0(t))] \times \vec{p} + \tau_v \frac{\hbar}{4m^*\Delta}(\nabla V_3) \times \vec{p} \quad (\text{1st - order VOI}).$$

The subscripts 'τ'/'0' here label valley –dependent/independent terms. $y_0(t)$ is explicitly written where the time-dependence appears. $H^{(1)}$ was previously derived and compared to the 1st-order relativistic correction (R.C.) in the standard Schrodinger equation of electrons. [24] For example, the 1st term in $H_0^{(1)}$ is the R.C. to the kinetic energy, and the 2nd and the 3rd terms in $H_0^{(1)}$ are the Darwin's term. The 2nd term in $H_0^{(1)}$ is a constant, and shall be dropped below, with no effect on the treatment. $H_\tau^{(1)}$ is the 1st-order valley-orbit interaction, the analogue of spin-orbit interaction. Similarly, for $H^{(2)}$, we write

$$H^{(2)} = H_0^{(2)} + H_\tau^{(2)}, \qquad (\text{III-1.2})$$

$$H_\tau^{(2)} = H_{\tau 2}^{(2)} + H_{\tau 3}^{(2)}, \quad (\text{2nd - order VOI})$$

$$H_{\tau 2}^{(2)} = -\tau_v \frac{3\hbar}{32m^{*2}\Delta^2}\left[\underline{\nabla V_2(x, y+y_0(t)) \times \vec{p}\vec{p}^2} + \vec{p}^2\,\nabla V_2(x, y+y_0(t)) \times \vec{p}\right]$$

$$H_{\tau 3}^{(2)} = -\tau_v \frac{3\hbar}{32m^{*2}\Delta^2}\left[\underline{\nabla V_3 \times \vec{p}\vec{p}^2} + \vec{p}^2\,\nabla V_3 \times \vec{p}\right]$$

$H_\tau^{(2)}$ is the 2nd-order VOI, and has been decomposed into $V_2$- and $V_3$- derived terms. Expressions underlined are evaluated first. $H_0^{(2)}$ is not given here, as it is irrelevant to the calculation of the VOI-based effect. See **Appendix B**. Note that the linearization of H(x,y,t) in $y_0$ leads to the approximation that H(x,y,t) ≈ H(x,y+$y_0$(t)), as can be verified with Eqns. (III-1), (III-1.1), and (III-1.2).

The analysis of VOI-based effect is carried out for the ground state, within the perturbation-theoretical framework where $H^{(0)}$ in Eqn. (III-1) is treated as the dominant term and $V_3$, $H^{(1)}$, and $H^{(2)}$ treated perturbatively. The conditions required for the perturbative calculation and the various energy scales involved are summarized below. First of all, $y_0 \ll Y$ and $\|H^{(2)}\| \ll \|H^{(1)}\| \ll \|H^{(0)}\|$, both of which have already appeared or been assumed above. We further take $\|V_3\| \ll \|H^{(0)}\|$ in order for $H^{(0)}$ to be the only dominant term in H. It leads to the following estimate of energy scales, namely, $\|H^{(0)}\| \sim E \sim \hbar w_0 \ll \Delta$, $\|H^{(1)}\| \sim O(\hbar^2 w_0^2/\Delta)$, and $\|H^{(2)}\| \sim O(\hbar^3 w_0^3/\Delta^2)$. We also assume the adiabatic condition, i.e., $w_s \ll w_0$, meaning that the AC field varies slowly in the time scale of the electron orbital motion. This permits us to employ the adiabatic perturbation theory to treat the time-dependence of H(t) due to the AC electric field. We note, in practical applications, that most of the above conditions can actually be relaxed, e.g., $\hbar w_0 \sim \Delta$, $\|H^{(2)}\| \sim \|H^{(1)}\| \sim \|H^{(0)}\|$, or $\|V_3\| \sim \|H^{(0)}\|$. In such cases, the Dirac equation and/or numerical work are required in an accurate analysis of the VOI-based effect, but the analytic calculation / result presented below can still serve as a useful guidance.

**Adiabatic perturbation-theoretical treatment**

We now perform the quantitative analysis of VOI-based effect. Specifically, we examine if the AC electric field is able to induce any valley dependence in the ground state of the QD-confined electron. We employ the adiabatic perturbation theory [25] and write the ground state wave function

$$\psi_0(x,y,t) \approx \varphi_0(x, y+y_0(t)) e^{-\frac{i}{\hbar}\int^t E_0(t')dt'} e^{\frac{i}{\hbar}\int^t \gamma_0(t')dt'}. \qquad (\text{III - 2})$$

Here, φ₀(t) and E₀(t) are the instantaneous ground state and energy, respectively, defined in the following,

$$H(x, y + y_0(t))\varphi_0(x, y + y_0(t)) = E_0(t)\varphi_0(x, y + y_0(t)). \quad (III - 2.1)$$

$\int^t E_0(t')dt'/\hbar$ and $\int^t \gamma_0(t')dt'/\int^t \gamma_0(t')dt'$ in (III-2) are, respectively, the dynamical phase and the geometric phase, of the state. Eqn. (III-2.1) has the following useful symmetry properties. Firstly, using the expression of H(t) provided in Eqns. (III-1), (III-1.1), and (III-1.2), one can show that $H(\tau_v = -1) = H^*(\tau_v = 1)$. Therefore, if φ₀(t) is an eigenstate solution in Eqn. (III-2.1) for $\tau_v = 1$, then $\varphi_0^*(t)$ is an eigenstate for $\tau_v = -1$, and both states are degenerate with the same energy E₀(t). Accordingly, the AC field does not generate any valley dependence in E₀(t), or in the dynamic phase. This is basically a consequence of the time-reversal symmetry for valley states. Secondly, in the case where V₃(x) is absent, we have $V_{QD}(x,y) = V_{QD}(-x,y)$ and, hence, $H(x,y) = H^*(-x,y)$. It follows that $\varphi_0(x,y) = \varphi_0^*(-x,y)$ as a result of the reflection symmetry. This fact shall be used later.

Substitution of the wave function (III-2) into the Schrodinger equation (III-1) yields γ₀, the rate of change in the geometric phase,

$$\gamma_0(t) = i\hbar < \varphi_0(x, y + y_0(t)) | \partial_t \varphi_0(x, y + y_0(t)) >$$
$$= -(\partial_t y_0(t)) < \varphi_0(x, y) | p_y \varphi_0(x, y) >. \quad (III - 2.2)$$

Note, in the 2ⁿᵈ line, that y₀(t) = 0 in <φ₀(τ_v)|p_yφ₀(τ_v)>. Therefore, we have the property i) γ₀ is linear in y₀. i) justifies the linearization of H in y₀, in the analysis of VOI-based effect. Using the time reversal property $\varphi_0(\tau_v = -1) = \varphi_0^*(\tau_v = 1)$ in (III-2.2) yields ii) <φ₀(τ_v)|p_yφ₀(τ_v)> α τ_v, or γ₀ α τ_v. ii) shows that, being valley dependent, γ₀ is able to generate

a valley-contrasting time evolution sought in the beginning of the section. Furthermore, in the case where $V_3$ is absent, the reflection property $\varphi_0(x,y) = \varphi_0^*(-x,y)$ mentioned earlier yields $\gamma_0 = 0$ in (III-2.2). This implies iii) $\gamma_0 \propto k_{3x}$. Collecting i), ii), and iii), we write

$$\gamma_0(t) \propto \tau_v [\partial_t y_0(t)] k_{3x}. \tag{III - 2.3}$$

This result serves as a useful guide in the evaluation of $\gamma_0$ below.

**$\gamma_0$ in the cubic case**

In (III-2.2), the adiabatic perturbative calculation has isolated the time dependence of $\gamma_0$, leaving only the time-independent expectation value, $<\varphi_0|p_y\varphi_0>|_{y_0=0}$, to be evaluated, with $\varphi_0$ now determined by the following (*time-independent*) equation

$$H|_{y_0=0} \varphi_0(x,y) = E_0 \varphi_0(x,y), \tag{III - 2.1'}$$
$$H = H^{(0)} + V_3 + H',$$
$$H' = H^{(1)} + H^{(2)}.$$

The notation H' is introduced above, and is to be treated within the time-independent perturbation theory in the evaluation of $<\varphi_0|p_y\varphi_0>|_{y_0=0}$. (From now on, the subscript $y_0=0$ shall be dropped.)

Utilizing the fact that $<\varphi_0|p_y\varphi_0> \propto \tau_v$, derived earlier, we write

$$<\varphi_0(x,y)|p_y\varphi_0(x,y)> \approx p_y^{(1)} + p_y^{(2,1)} + p_y^{(2,2)}, \tag{III - 3}$$

correct to the 2$^{nd}$-order R.C. $p_y^{(1)}$ is the 1$^{st}$-order R.C. and derived from $H_\tau^{(1)}$ in the 1$^{st}$-order perturbative treatment of H'. $p_y^{(2,1)}$ denotes the 2$^{nd}$- order R.C., derived from $H_\tau^{(1)}$, in the 2$^{nd}$-order perturbative treatment of H'. $p_y^{(2,2)}$ denotes the 2$^{nd}$- order R.C., derived from $H_\tau^{(2)}$, in the 1$^{st}$-order perturbative treatment of H'. We summarize these perturbative results below:

$$p_y^{(1)} = 0, \qquad (\text{III - 3.1})$$

$$p_y^{(2,1)} = -\tau_v \frac{s}{64}\left[-2 + \frac{4}{s} - \frac{16}{3(s+1)} + \frac{50}{3(s+2)}\right]\hbar\frac{\hbar^2 w_0^2}{\Delta^2} k_{3x},$$

$$p_y^{(2,2)} = \tau_v \frac{1}{16}\hbar\frac{\hbar^2 w_0^2}{\Delta^2} k_{3x}.$$

For details of the derivation, see **Appendix B**. Overall, this gives

$$\gamma_0(t) = C_3 \tau_v \hbar (\partial_t y_0)\frac{\hbar^2 w_0^2}{\Delta^2} k_{3x}, \qquad (\text{III - 4})$$

$$C_3 = \frac{s}{96}\left[-3 - \frac{8}{s+1} + \frac{25}{s+2}\right].$$

We stress that because $p_y^{(1)} = 0$, an analysis correct only to the 1$^{st}$-order R.C. here would have yielded a vanishing $\gamma_0$. The finite result shown in Eqn. (III-4) is basically a *2$^{nd}$-order relativistic type effect* involving the VOI ($H_\tau^{(1)}$ and $H_\tau^{(2)}$).

**$\gamma_0$ in the quartic case**

We extend the result of (III-4) to the quartic case where $V_{QD} = V_2(x,y) + e\,\varepsilon_x\,x + V_4(x)$, with $V_4(x) = 1/4\ m^* w_0^2\ k_{4x} x^4$. For $k_{4x} < 0$, the potential $V_2(x,y) + V_4(x)$ is a realistic description of a finite, symmetric confinement potential in the x-direction.

In the limit of weak $\varepsilon_x$ (with $x_0 \ll Y$), expansion of $V_{QD}$ around $x_0$ leads to

$$V_{QD} = V_2(x,y) + e\varepsilon_x x + V_4(x)$$
$$\stackrel{x'=x+x_0}{\approx} V_2(x',y) + V_4(x',y) - m^* w_0^2 k_{4x} x_0 x'^3.$$

Here, nonlinear terms of $x_0$ have been dropped. The result shows that the electric field produces effectively a cubic term in the potential with the strength $k_{3x} = -3 k_{4x} x_0$. In the limit where $|V_4| \ll V_2$, we substitute the effective $k_{3x}$ into (III- 4), yielding

$$\gamma_0(t) = C_4(s)\tau_v \hbar(\partial_t y_0)\frac{\hbar^2 w_0^2}{\Delta^2} k_{4x} x_0, \qquad (III-5)$$

$$C_4(s) = -\frac{s}{32}\left[-3 - \frac{8}{s+1} + \frac{25}{s+2}\right].$$

$C_4$ can be optimized, by a choice of the parameter s (i.e., the QD shape). We briefly note the following. In the case of an isotropic potential (s = 1), $C_4$ = -1/24. In the anisotropic case, $C_4$ varies slowly with s, with

$$|C_4| > \frac{1}{24} \text{ for } s < 1, \quad |C_4| < \frac{1}{24} \text{ for } s > 1.$$

In practical applications, the conditions, $x_0 \ll Y$ and $|V_4| \ll V_2$, used in the derivation of Eqn. (III-5), can be relaxed, e.g., $x_0 \sim Y$ and $|V_4| \sim V_2$.

**Qubit manipulation**

For illustration, we consider the qubit manipulation in the quartic case. As made clear in the above, a geometric phase contrast is induced by the AC electric field between valley states. In half of the AC cycle ($-\pi/2w_s$, $\pi/2w_s$), for example, it evolves as follows,

$$|K> \to |K> e^{i\Phi_{1/2}}, \tag{III-6}$$

$$|K'> \to |K'> e^{-i\Phi_{1/2}},$$

$$\Phi_{1/2} = \frac{1}{\hbar} \int_{-\pi/2w_s}^{\pi/2w_s} \gamma_0(t'; \tau_v = 1) dt' = \frac{y_0^{(\max)}}{l_{vo}},$$

$$y_0^{(\max)} = \frac{e\varepsilon_y}{m^* w_0^2}, \quad l_{vo} = \left(2C_4 \frac{\hbar^2 w_0^2}{\Delta^2} k_{4x} x_0\right)^{-1}.$$

Here, $y_0^{(\max)}$ is the electron displacement amplitude due to the AC field, and $l_{vo}$ is called the valley-orbit length.

In the case of a valley pair qubit, the foregoing phase contrast results in a qubit state transformation, as follows. Let us consider the simple mode of manipulation where the exchange interaction J between the QDs is turned down when both QDs (or just one of them) are subject to AC electric fields. For $J/\hbar \ll w_s$, it freezes the inter-dot orbital motion, $|K_L K'_R> \leftrightarrow |K'_L K_R>$, and hence the associated J-induced $R_z$ during the action of the AC fields. Based on (III-6), the AC fields produce the following evolution of valley pair states in half of the AC cycle,

$$\begin{array}{l}|K_L K_R'> \to e^{i\theta_x/2} |K_L K_R'> \\ |K_L' K_R> \to e^{-i\theta_x/2} |K_L' K_R>\end{array} \quad or \quad \begin{array}{l}|x_+> \to e^{i\theta_x/2} |x_+> \\ |x_-> \to e^{-i\theta_x/2} |x_->\end{array},$$

$$\frac{\theta_x}{2} = \frac{y_{0,L}^{(\max)}}{l_{vo,L}} - \frac{y_{0,R}^{(\max)}}{l_{vo,R}}. \tag{III-7}$$

Here, $y_{0,L(R)}^{(\max)}$ is the AC field-induced electron displacement amplitude in $QD_{L(R)}$, and $l_{vo,L(R)}$ is the valley-orbit length for $QD_{L(R)}$. (III-7) represents a state rotation about the x-axis, $R_x(\theta_x)$. Combining $R_x(\theta_x)$ and $R_z(\theta_z)$, one can manipulate the qubit to an arbitrary point on the Bloch sphere. See Fig. 2.

We give below a numerical estimate of the time (denoted as $t_{operation}$) needed for a typical single qubit manipulation. It is assumed that a series of alternate $R_x(\theta_x)$'s and $R_z(\theta_z = \pi)$'s are used in the manipulation, as shown in Fig. 2. We take $\pi/w_s \sim 0.1$ ns. Moreover, J in the range of 1meV or lower is achievable.[5] Therefore, the time ($\pi\hbar/J$) spent on each $R_z(\theta_z = \pi)$ can be made much less than or comparable to the time ($\pi/w_s$) on each $R_x(\theta_x)$. Accordingly, $t_{operation}$ is determined primarily by the total time spent on $R_x(\theta_x)$'s, and it leads to the estimate that $t_{operation} \sim O[\pi/w_s\theta_x]$. Using $k_{4x} = L^{-2}$ (L = QD size), $x_0 = 0.3L$, s = 1, $\hbar w_0/\Delta = 0.5$, Eqn. (III-6) gives $l_{vo,L(R)} \sim 120L$. For $y_{0,L}^{(max)} = - y_{0,R}^{(max)} = 0.3L$, and $\pi/w_s = 0.1$ ns, Eqn. (III-7) gives $\theta_x = 0.01$ and, thus, $t_{operation} \sim O(10$ ns$)$.

## IV. Optical response and quantum state transfer

Firstly, we describe the near-band-gap optical response from a gapped graphene QD. In particular, we consider the excitation of an electron from a valence band state to the lowest quantized conduction band state in the QD, as shown in Fig. 3.

Since the excitation involves both valence and conduction bands, we return to the two-band model, i.e., the Dirac equation,

$$(H_D^{(0)} + H_A)\phi_D = i\hbar\partial_t\phi_D, \qquad (IV-1)$$

$$H_D^{(0)} = \begin{pmatrix} \Delta + V_{QD} & v_F\hat{p}_- \\ v_F\hat{p}_+ & -\Delta + V_{QD} \end{pmatrix}, H_A = \begin{pmatrix} 0 & ev_F A_- \\ ev_F A_+ & 0 \end{pmatrix},$$

$$\phi_D = \begin{pmatrix} \varphi_A \\ \varphi_B \end{pmatrix},$$

$$\hat{p}_- = p_x - i\tau_v p_y, \hat{p}_+ = p_x + i\tau_v p_y,$$

$$A_- = A_x - i\tau_v A_y, A_+ = A_x + i\tau_v A_y.$$

Here, **A** = ($A_x$, $A_y$) is the vector potential of the radiation field. $H_D^{(0)}$ is the QD Hamiltonian in the absence of radiation, and $H_A$ is the light-electron interaction. We describe briefly the eigenstates of $H_D^{(0)}$ below. We denote $\Phi_D^{(0,c)} = (\varphi_A^{(0,c)}, \varphi_B^{(0,c)})^T$ and $\Phi_D^{(0,v)} = (\varphi_A^{(0,v)}, \varphi_B^{(0,v)})^T$ (T = transpose) as the lowest quantized conduction band state in the QD and the near-band-edge valence band state around the QD, respectively, with $E_0^{(c)}$ and $E_0^{(v)}$ being the corresponding energies. From Eqn. (IV-1),

$$(\varphi_A^{(0,c)}, \varphi_B^{(0,c)})|_{\tau_v = -1} = (\varphi_A^{(0,c)*}, -\varphi_B^{(0,c)*})|_{\tau_v = 1}, \qquad (\text{IV-2a})$$

$$(\varphi_A^{(0,v)}, \varphi_B^{(0,v)})|_{\tau_v = -1} = (\varphi_A^{(0,v)*}, -\varphi_B^{(0,v)*})|_{\tau_v = 1},$$

due to the time reversal symmetry between valley states in the absence of radiation. Moreover, for electrons near the gap, $|\varphi_A^{(0,c)}| \gg |\varphi_B^{(0,c)}|$ and $|\varphi_A^{(0,v)}| \ll |\varphi_B^{(0,v)}|$ for both $\tau_v = \pm 1$. In fact, it can be verified that

$$|\varphi_B^{(0,c)}| / |\varphi_A^{(0,c)}| \sim |\varphi_A^{(0,v)}| / |\varphi_B^{(0,v)}| \sim O(E/\Delta)^{1/2}. \qquad (\text{IV-2b})$$

We consider the near-resonance optical response of the QD to a normally incident light, in the form of a circularly polarized (σ+ or σ-) plane wave. The light-electron interaction in the case is given in the following

$$H_A\Big|_{(\tau_v=1,\sigma+)\,or\,(\tau_v=-1,\sigma-)} = ev_F A_0 \left[ e^{-i(k_{ph}z - w_{ph}t)} \begin{pmatrix} 0 & 1 \\ 0 & 0 \end{pmatrix} + e^{i(k_{ph}z - w_{ph}t)} \begin{pmatrix} 0 & 0 \\ 1 & 0 \end{pmatrix} \right],$$

$$H_A\Big|_{(\tau_v=-1,\sigma+)\,or\,(\tau_v=1,\sigma-)} = ev_F A_0 \left[ e^{i(k_{ph}z - w_{ph}t)} \begin{pmatrix} 0 & 1 \\ 0 & 0 \end{pmatrix} + e^{-i(k_{ph}z - w_{ph}t)} \begin{pmatrix} 0 & 0 \\ 1 & 0 \end{pmatrix} \right],$$

where $k_{ph}$ = photon wave vector, $w_{ph}$ = photon frequency, and $A_0$ is the amplitude of **A**. $H_A$ is treated with the time-dependent perturbation theory. We take z = 0 in the graphene plane.

Then, near resonance ($\hbar w_{ph} \sim E_0^{(c)} - E_0^{(v)}$), the optical response is governed by the following optical matrix elements

$$M_> = ev_F A_0 <\varphi_A^{(c)} | \varphi_B^{(v)}>, \quad \text{for } (\tau_v = 1, \sigma_+) \text{ or } (\tau_v = -1, \sigma_-), \qquad \text{(IV-3)}$$
$$M_< = ev_F A_0 <\varphi_B^{(c)} | \varphi_A^{(v)}>, \quad \text{for } (\tau_v = 1, \sigma_-) \text{ or } (\tau_v = -1, \sigma_+).$$

From the properties of wave functions listed in (IV-2a) and (IV-2b), we obtain

$$M_>(\tau_v = 1) = - M_>(\tau_v = -1)^*, \quad M_<(\tau_v = 1) = - M_<(\tau_v = -1)^*, \qquad \text{(IV-4)}$$

$$|M_<| / |M_>| \sim O(E/\Delta). \qquad \text{(IV-5)}$$

Eqn. (IV-5) permits us to make the approximation that $M_<(\tau_v = \pm 1) = 0$, leading to the approximate selection rule that |K(valence)> + |σ+> → |K(conduction)> and |K'(valence)> + |σ-> → |K'(conduction)>, mentioned previously in **Sec. I** and plotted in Fig. 3. Moreover, although derived for the band-to-band transition, the above result (Eqns. (IV-3) ~ (IV-5)) applies to excitonic excitations as well, except for an additional enhancement factor in the matrix element in (IV-3), due to the sizable overlap of electron and hole wave functions in the exciton state.[26] Below, we denote $M_>(\tau_v = 1)$ simply as $M_>$, and $M_<(\tau_v = 1)$ as $M_<$.

Next, we describe a feasible method to transfer the quantum state from a photon qubit to a valley pair qubit, shown in Fig. 4, based on the optical response described in Eqns. (IV-3) ~ (IV-5) for the QD. To begin, the valley pair qubit is placed in a photonic cavity, and initialized in the singlet state $|K_L K'_R> - |K'_L K_R>$, with the exchange coupling J turned down after the initialization, in order to freeze the inter-QD orbital motion during the QST that follows. We take that the energy levels in $QD_L$ and $QD_R$ are detuned (by back gate voltages), and the cavity photon energy ($\hbar w_{cavity}$) matches only the exciton binding energy ($\hbar w_{exciton}$) in $QD_L$. As shown in Fig. 4, the photon signal (with the photon frequency $w_{ph} = w_{cavity}$) enters the cavity in the polarization state $\alpha|\sigma+\rangle+\beta|\sigma-\rangle$ carrying the quantum information, and interact with the electrons in the QDs. Then, due to the light-electron interaction, the initial (photon-electron composite) state evolves in time. According to Eqns. (IV-3) ~ (IV-5), the following cavity QED processes take place in $QD_L$,

photon absorption:

$|\Phi_0\rangle = (|K_L K'_R\rangle - |K'_L K_R\rangle) \times (\alpha|\sigma+\rangle+\beta|\sigma-\rangle) \rightarrow$

$|\Phi_1\rangle = (-\beta M_>^* + \alpha M_<) |K'_{ex,L} K_L K'_R\rangle - (\alpha M_> - \beta M_<^*) |K_{ex,L} K'_L K_R\rangle$,

photon re-emission ($\rightarrow$) and re-absorption ($\leftarrow$):

$|\Phi_1\rangle \leftrightarrow$

$|\Phi_2\rangle = (-\beta M_>^* + \alpha M_<) [-M_> |\sigma- K_L K'_R\rangle + M_<^* |\sigma+ K_L K'_R\rangle] -$

$(\alpha M_> - \beta M_<^*) [M_>^* |\sigma+ K'_L K_R\rangle - M_< |\sigma- K'_L K_R\rangle]$,

where $|\Phi_0\rangle$ is the initial state, $K_{ex,L}$ ($K'_{ex,L}$) is the K (K')-valley exciton created by the process of photon absorption in $QD_L$. Note that the intermediate state $|\Phi_2\rangle$ generated in the above processes is entangled and the photon state can no longer be factored out in $|\Phi_2\rangle$. In this entangled state, the information carried by the photon is shared between photons and valley electrons. Eventually, the re-emitted photon in $|\Phi_2\rangle$ escapes the cavity, and can be measured for its state of linear polarization, leaving the valley pair qubit carrying the full information. The measurement produces the valley pair state

$|\Phi_{3x}\rangle = \langle\sigma_x|\Phi_2\rangle$

$= (-\beta M_>^* + \alpha M_<)(-M_> + M_<^*) |K_L K'_R\rangle - (\alpha M_> - \beta M_<^*)(M_>^* - M_<) |K'_L K_R\rangle,$ (IV-6a)

if x-polarization is measured, or the state

$|\Phi_{3y}\rangle = \langle\sigma_y|\Phi_2\rangle$

$= (-\beta M_>^* + \alpha M_<)(M_> + M_<^*) |K_L K'_R\rangle - (\alpha M_> - \beta M_<^*)(M_>^* + M_<) |K'_L K_R\rangle,$ (IV-6b)

if it yields y-polarization.

In order to see that the above procedure indeed produces the desired photon → valley QST, we make the approximation $M_< = 0$ in (IV-6a) and (IV-6b). Then, we see that

$|\Phi_{3x}\rangle \approx \beta |K_L K'_R\rangle - \alpha |K'_L K_R\rangle,$ (IV-6a')

$|\Phi_{3y}\rangle \approx \beta |K_L K'_R\rangle + \alpha |K'_L K_R\rangle.$ (IV-6b')

Thus, the quantum information is successfully transferred to the valley pair qubit. If desired, one can further manipulate the qubit state in (IV-6a') or (IV-6b') into the state $\alpha |K_L K'_R\rangle + \beta |K'_L K_R\rangle$ or $\alpha|z_S\rangle + \beta|z_{T0}\rangle$, with the VOI-based method described in **Sec. III**.

It is noted, in the QST mechanism envisioned above, that some information distortion (of $O(E/\Delta)$) appears to occur in the QST, as reflected in the contrast between Eqns. (IV-6a), (IV-6b) and Eqns. (IV-6a'), (IV-6b'), due to the finite magnitude of $M_<$. However, the distortion would only turn into a true loss of fidelity, if the transfer stands alone not being a part of a series of QSTs. As shall be shown below, in the back-to-back valley → photon → valley QST, the distortion of information in one transfer is cancelled by that in the other.

Let us now consider the valley → photon QST. We start with the following valley pair state in the cavity, $\alpha|K_L K'_R\rangle + \beta|K'_L K_R\rangle$, carrying quantum information. We send a photon into the cavity, with the photon initialized in the state $|\sigma+\rangle+|\sigma-\rangle$, and let it interact with the electrons. The photon-electron state evolves as follows,

photon absorption:

$|\Phi_0'\rangle = (\alpha|K_L K'_R\rangle + \beta|K'_L K_R\rangle) \times (|\sigma+\rangle+|\sigma-\rangle) \rightarrow$

$|\Phi_1'\rangle = \alpha (M_< - M_>^*) |K'_{ex,L} K_L K'_R\rangle + \beta (M_> - M_<^*) |K_{ex,L} K'_L K_R\rangle$,

photon re-emission (→) and re-absorption (←):

$|\Phi_1'\rangle \leftrightarrow$

$|\Phi_2'\rangle = \alpha (M_< - M_>^*) [-M_> |\sigma- K_L K'_R\rangle + M_<^* |\sigma+ K_L K'_R\rangle] +$

$\beta (M_> - M_<^*) [M_>^* |\sigma+ K'_L K_R\rangle - M_< |\sigma- K'_L K_R\rangle]$.

In $|\Phi_2'\rangle$, the re-emitted photon and the QD electrons are entangled. The re-emitted photon eventually escapes from the cavity, leaving behind the valley pair. We measure the valley pair state, producing the following photon state

$|\Phi_{3S}'\rangle = \langle z_S|\Phi_2'\rangle$

$= [\alpha M_<^* (M_< - M_>^*) - \beta M_>^* (M_> - M_<^*)] |\sigma+\rangle +$

$[-\alpha M_> (M_< - M_>^*) + \beta M_< (M_> - M_<^*)] |\sigma-\rangle$, (IV-7a)

if the measurement yields the singlet state, or

$|\Phi_{3T0}'\rangle = \langle z_{T0}|\Phi_2'\rangle$

$= [\alpha M_<^* (M_< - M_>^*) + \beta M_>^* (M_> - M_<^*)] |\sigma+\rangle -$

$[\alpha M_> (M_< - M_>^*) + \beta M_< (M_> - M_<^*)] |\sigma-\rangle,$ (IV-7b)

if the triplet state is measured.

We remark on two points. Firstly, if we set $M_< = 0$ in (IV-7a) and (IV-7b), we obtain

$|\Phi_{3S}'\rangle = -\beta|\sigma+\rangle + \alpha|\sigma-\rangle,$ (IV-7a')

$|\Phi_{3T0}'\rangle = \beta|\sigma+\rangle + \alpha|\sigma-\rangle,$ (IV-7b')

showing a successful valley → photon QST. Again, because of the finite magnitude of $M_<$, the quantum information appears to be distorted in the QST, and would cause true fidelity loss of $O(E/\Delta)$, if the transfer is not coupled with other QSTs. Secondly, if we combine the results of Eqns. (IV-6a), (IV-6b), (IV-7a), and (IV-7b), it can be shown that the back-to-back QST is highly faithful, as expressed in the following diagram,

$$\alpha |K_L K'_R\rangle + \beta |K'_L K_R\rangle \qquad\qquad\qquad (IV-8)$$
$$\xrightarrow{valley\ to\ photon} \alpha'|\sigma+\rangle + \beta'|\sigma-\rangle$$
$$\xrightarrow{photon\ manipulation} -\alpha'|\sigma+\rangle + \beta'|\sigma-\rangle$$
$$\xrightarrow{photon\ to\ valley} \alpha |K_L K'_R\rangle + \beta |K'_L K_R\rangle.$$

Here, $\alpha' = \alpha M_<^* e^{i\chi} - \beta M_>^*$, $\beta' = -\alpha M_> e^{i\chi} + \beta M_<$, and $e^{i\chi} = (M_< - M_>^*)/(M_> - M_<^*)$. In the diagram, we have assumed that the singlet state is measured in the valley → photon QST and that the x-polarization is measured in the photon → valley QST. Faithful QST can also be shown with different results of measurement.

It is worth noting that, given the highly faithful back-to-back process shown above, a similar but longer process such as valley → photon → ……→ valley, which involves valley ↔ photon QST for many times, is obviously, in principle, as faithful as the back-to-back process. That is, the small quantum distortion occurring in the single step valley ↔ photon QST does not accumulate along the way. This is an important feature of the present valley-based approach for quantum memories, and is also an essential requirement for any quantum memories employed in long distance QCs. In reality, there are various factors which affect the yield and fidelity in the QST envisioned here, such as cavity Q-factor and valley state decoherence. These important issues shall be studied in a separate work.

**V. Summary and conclusion**

In summary, we have investigated valley pair qubits in graphene double quantum dots, in the in-plane magnetic field configuration, and developed a method of qubit manipulation for this configuration. The method is based on the $2^{nd}$-order relativistic type effect in gapped graphene involving the valley-orbit interaction, and is able to operate in the time scale of O(10ns). Moreover, the work has also considered the optical response of graphene quantum dots, in terms of valley excitation with respect to photonic polarization, and illustrated faithful quantum state transfers from photon to valley pair qubits and vice versa. It shows the potential of graphene quantum dots in photonic quantum communications, and in particular, the feasibility of implementing graphene-based quantum memories for quantum repeaters. Along with the previous exploration in Reference 5 of valley pair qubits for quantum computing, it suggests the interesting prospect of an all-graphene approach for the solid state components of a quantum network, i.e., quantum computers and quantum memories in communications.

**Acknowledgement** - We thank the support of ROC National Science Council through the contract No. NSC100-2112-M-007-009.

**Appendix A**

**The Schrodinger type equation in gapped graphene, including the 2$^{nd}$-order R.C.**

We derive the Schrodinger type equation including the 2$^{nd}$-order R.C., for near-band-edge electrons/holes in gapped graphene. Below, we consider only the case of electrons. (The case of holes can be worked out analogously.) We begin with the Dirac type equation in the two-band model,

$$\begin{pmatrix} V - E & v_F \hat{p}_- \\ v_F \hat{p}_+ & -2\Delta + V - E \end{pmatrix} \begin{pmatrix} \varphi_A \\ \varphi_B \end{pmatrix} = 0.$$

Here, V = potential energy, E = electron energy with respect to the conduction band edge, and $p_\pm = p_x \pm i p_y$. Or, equivalently,

$$\varphi_B = \frac{1}{2\Delta + E - V} v_F \hat{p}_+ \varphi_A. \tag{A-1a}$$

$$H' \varphi_A = E \varphi_A, \tag{A-1}$$

$$H' = v_F \hat{p}_- \left( \frac{1}{2\Delta + E - V} v_F \hat{p}_+ \right) + V,$$

(A-1a) is the 1$^{st}$-order differential equation corresponding to the 2$^{nd}$ row of the Dirac equation. (A-1) is a 2$^{nd}$-order differential equation obtained by combining the two 1$^{st}$-order differential equations in the Dirac equation, and is the primitive form of the Schrodinger equation to be derived.

We expand the energy denominator in (A-1), up to the order $O(E^2/\Delta^2)$, giving

$$H' \approx v_F \hat{p}_- \left( \frac{1}{2\Delta}[1 - \frac{E-V}{2\Delta} + (\frac{E-V}{2\Delta})^2 ]v_F \hat{p}_+ \right) + V \quad \text{(A-2)}$$

$$= H^{(0)} + H^{(1)\prime} + H^{(2)\prime}.$$

Here,

$$H^{(0)} = p^2/2m^* + V, \ (m^* = \Delta/v_F^2) \quad \text{(A-3)}$$

deriving from the 1$^{st}$ term in […] (and V), and constituting the "non-relativistic" part of the "Schrodinger Hamiltonian". $H^{(1)\prime}$ derives from the 2$^{nd}$ term in […], and constitutes the 1$^{st}$-order R.C. It was already given previously [5], and listed below,

$$H^{(1)\prime} = -\frac{1}{4m^*\Delta} \underline{\vec{p}V} \cdot \vec{p} - \frac{1}{4m^*\Delta} \underline{\vec{p}^2 V} + \tau_v \frac{\hbar}{4m^*\Delta} \underline{\nabla V} \times \vec{p}. \quad \text{(A-4)}$$

Here, terms underlined are evaluated first. We focus on the derivation of $H^{(2)\prime}$ (the 2$^{nd}$-order R.C.) below.

There are two contributions to $H^{(2)\prime}$, with one coming from the 2$^{nd}$ term in […] in (A-2), and the other from the 3$^{rd}$ term in […]. We denote them as $H_2^{(2)\prime}$ and $H_3^{(2)\prime}$, respectively. We have

$$H^{(2)\prime} = H_2^{(2)\prime} + H_3^{(2)\prime}, \quad \text{(A-5)}$$

$$H_2^{(2)\prime}\varphi_A = -\frac{1}{2m^*} \hat{p}_- \left( \frac{E-V}{2\Delta} \right) \hat{p}_+ \varphi_A \Big|_{\text{2nd order}}$$

$$= \frac{1}{32m^{*3}\Delta^2} \vec{p}^6 \varphi_A + \frac{1}{16m^{*2}\Delta^2} \vec{p}^2 \underline{\vec{p}^2 V}\varphi_A + \frac{1}{16m^{*2}\Delta^2} \vec{p}^2 \underline{\vec{p}V} \cdot \vec{p}\varphi_A$$

$$- \tau_v \frac{\hbar}{16m^{*2}\Delta^2} \vec{p}^2 \underline{\nabla V \times \vec{p}}\varphi_A, \quad \text{(A-5.1)}$$

and

$$H_3^{(2)'}\varphi_A = \frac{1}{8m^{*}\Delta^2}\hat{p}_-\left((E-V)^2\hat{p}_+\varphi_A\right)$$

$$= R_1^{(2)} + R_2^{(2)},$$

$$R_1^{(2)} = \frac{1}{16m^{*2}\Delta^2}\left[\frac{1}{2m^{*}}\vec{p}^{\,4}\varphi_A + 2\underline{\vec{p}V}\cdot\underline{\vec{p}\varphi_A} + \underline{\vec{p}^{\,2}V\varphi_A}\right],$$

$$R_2^{(2)} = \frac{1}{8m^{*2}\Delta^2}\left[\underline{\vec{p}V}\cdot\underline{\vec{p}\vec{p}^{\,2}\varphi_A} + \underline{\vec{p}^{\,2}V}\cdot\underline{\vec{p}^{\,2}\varphi_A} - \tau_v\hbar\underline{\nabla V}\times\underline{\vec{p}\vec{p}^{\,2}\varphi_A}\right] \qquad (A\text{-}5.2)$$

With the results of (A-3), (A-4), (A-5.1), and (A-5.2), it would appear that we have finished the derivation of the Schrodinger equation. However, it can be verified that several terms of H' as given in Eqns. (A-4), (A-5a), and (A-5b) are not Hermitian and, hence, H' is not Hermitian. This is tied to the fact that the wave function $\varphi_A$ used with the above Schodinger equation is not normalized, being just one of the two components in the Dirac wave function. This can be rectified by a similarity transformation, as follows.

We introduce the following transformation,

$$\varphi_A \to \psi = \Omega\varphi_A, \quad H' \to H = \Omega H'\Omega^{-1},$$

$$\Omega = 1 + \Omega^{(1)} + \Omega^{(2)},$$

$$\Omega^{(1)} = \frac{1}{8m^{*}\Delta}\vec{p}^{\,2},$$

$$\Omega^{(2)} = \frac{-9}{128m^{*2}\Delta^2}\vec{p}^{\,4} + \tau_v\frac{\hbar}{8m^{*}\Delta^2}\underline{\nabla V}\times\vec{p} - \frac{1}{16m^{*}\Delta^2}\underline{\vec{p}^{\,2}V}.$$

It can be shown that the transformed wave function ψ is normalized to the 2$^{nd}$-order R.C. With the above transformation, we obtain the Schrodinger euqation,

$$H\psi = E\psi, \tag{A-6}$$
$$H = H^{(0)} + H^{(1)} + H^{(2)},$$
$$H^{(1)} = -\frac{\vec{p}^4}{8m^{*2}\Delta} - \frac{1}{8m^*\Delta}\vec{p}^2 V + \tau_v \frac{\hbar}{4m^*\Delta}\nabla V \times \vec{p},$$
$$H^{(2)} = \frac{\vec{p}^6}{16m^{*3}\Delta} - \frac{1}{64m^{*2}\Delta^2}\vec{p}^2 V \vec{p}^2 + \frac{3}{64m^{*2}\Delta^2}\{\vec{p}^2 V, \vec{p}^2\}_+$$
$$+ \frac{1}{128m^{*2}\Delta^2}\{\vec{p}^4, V\}_+ - \tau_v \frac{3\hbar}{32m^{*2}\Delta^2}\{\nabla V \times \vec{p}, \vec{p}^2\}_+.$$

In (A-6), it is easy to verify that H$^{(0)}$ and H$^{(1)}$ are both Hermitian. Moreover, each {…}$_+$ in H$^{(2)}$ is the symmetrized product of two Hermitian operators and, consequently, H$^{(2)}$ is Hermitian. Altogether, it gives a Hermitian Hamiltonian H.

**Appendix B**

**Geometric phase rate of change (γ₀) due to the ac electric field-induced, VOI-based effect**

We provide the perturbative evaluation of $\gamma_0$ due to the ac electric field-induced, VOI-based effect. According to **Sec. III**, the rate of change in the geometric phase is

$$\gamma_0(t) = -(\partial_t y_0(t)) < \varphi_0(x,y) \mid p_y \varphi_0(x,y) >, \qquad (\text{III - 2.2})$$

with $y_0(t) = 0$ in $<\varphi_0|p_y\varphi_0>$. Here, $y_0(t) = \dfrac{e\varepsilon_y \sin w_s t}{m^* w_0^2}$ is the ac field-induced displacement, and $\varphi_0$ is determined by the following *time-independent* Schrodinger type equation including up to the 2$^{nd}$-order R.C., derived in **Appendix A** and presented in **Sec. III**,

$$H\big|_{y_0=0} \varphi_0(x,y) = E_0 \varphi_0(x,y), \qquad (\text{III - 2.1'})$$
$$H = H^{(0)} + V_3(x) + H',$$
$$H' = H^{(1)} + H^{(2)},$$
$$V_3(x) = \frac{1}{3} k_{3x} m^* w_0^2 x^3,$$

$$H^{(0)}(x,y,t) = \frac{\vec{p}^2}{2m^*} + V_2(x,y), \qquad (\text{III -1})$$
$$V_2(x,y) = \frac{1}{2} m^* w_0^2 (x^2 + s^2 y^2),$$

$$H^{(1)} = H_0^{(1)} + H_\tau^{(1)}, \tag{III-1.1}$$

$$H_\tau^{(1)} = \tau_v \frac{\hbar}{4m^{*}\Delta}[\nabla(V_2 + V_3)] \times \vec{p} \quad \text{(1st - order VOI)},$$

$$H_0^{(1)} = -\frac{\vec{p}^4}{8m^{*2}\Delta} - \frac{1}{8m^{*}\Delta}(\vec{p}^2 V_2) - \frac{1}{8m^{*}\Delta}(\vec{p}^2 V_3)$$

$$\rightarrow H_{02}^{(1)} + H_{03}^{(1)}$$

$$H_{02}^{(1)} = -\frac{p^4}{8m^{*2}\Delta}, \quad H_{03}^{(1)} = -\frac{1}{8m^{*}\Delta}(\vec{p}^2 V_3),$$

$$H^{(2)} = H_0^{(2)} + H_\tau^{(2)}, \tag{III-1.2}$$

$$H_\tau^{(2)} = H_{\tau 2}^{(2)} + H_{\tau 3}^{(2)}, \quad \text{(2nd - order VOI)}$$

$$H_{\tau 2}^{(2)} = -\tau_v \frac{3\hbar}{32m^{*2}\Delta^2}\left[\nabla V_2 \times \vec{p}\vec{p}^2 + \vec{p}^2 \nabla V_2 \times \vec{p}\right],$$

$$H_{\tau 3}^{(2)} = -\tau_v \frac{3\hbar}{32m^{*2}\Delta^2}\left[\nabla V_3 \times \vec{p}\vec{p}^2 + \vec{p}^2 \nabla V_3 \times \vec{p}\right]$$

Note, in (III-1.1), that the term $\frac{1}{8m^{*}\Delta}(\vec{p}^2 V_2)$ in $H_0^{(1)}$ has been dropped, being only a constant.

According to **Sec. III**, we write

$$<\varphi_0(x,y) | p_y \varphi_0(x,y)> \approx p_y^{(1)} + p_y^{(2,1)} + p_y^{(2,2)} \tag{III-3}$$
$$\propto \tau_v k_{3x},$$

correct to the $2^{nd}$-order R.C., where $p_y^{(1)}$ is the $1^{st}$-order R.C. and derives from $H_\tau^{(1)}$ in the $1^{st}$-order perturbative treatment of H'. $p_y^{(2,1)}$ denotes the $2^{nd}$- order R.C., deriving from $H_\tau^{(1)}$, in the $2^{nd}$-order perturbative treatment of H'. $p_y^{(2,2)}$ denotes the $2^{nd}$- order R.C., deriving from $H_\tau^{(2)}$, in the $1^{st}$-order perturbative treatment of H'. We calculate the three foregoing momentum matrix elements below.

Firstly, we show that $p_y^{(1)} = 0$. Because $p_y^{(1)}$ is $1^{st}$ order in R.C., it simplifies by writing

$$p_y^{(1)} = <\varphi_0'| p_y \varphi_0'>,$$
$$(H^{(0)} + V_3 + H_0^{(1)} + H_\tau^{(1)})\varphi_0' = E_0'\varphi_0',$$

where $\varphi_0$ in $p_y^{(1)}$ has been replaced by $\varphi_0'$, the ground state solution correct only to the 1$^{st}$-order R.C. Note that it suffices to evaluate $\varphi_0'$ here with the 1$^{st}$-order perturbation theory treating $H_0^{(1)}$ and $H_\tau^{(1)}$ as the perturbation.

Since $p_y^{(1)} \propto \tau_v$, only the perturbative correction due to $H_\tau^{(1)}$ contributes to $p_y^{(1)}$. We thus drop $H_0^{(1)}$ in the Schrodinger equation and rewrite

$$p_y^{(1)} = <\varphi_0'| p_y \varphi_0'>,$$
$$(H^{(0)} + V_3 + H_\tau^{(1)})\varphi_0' = E_0'\varphi_0'.$$

(The same notation $\varphi_0'$ shall be used repeatedly where it causes no confusion.) Applying Ehrenfest's theorem to the following expectation value involving the last $\varphi_0'$,

$$0 = \frac{d <\varphi_0'| y |\varphi_0'>}{dt} = -\frac{i}{\hbar} <\varphi_0'|[y, H^{(0)} + V_3 + H_\tau^{(1)}]|\varphi_0'>,$$

we obtain

$$p_y^{(1)} = -\tau_v \frac{\hbar}{4\Delta} <\varphi_0'| \partial_x V_{23} |\varphi_0'>,$$

where $V_{23} = V_2 + V_3$. The above procedure has extracted out an order of R.C. and provided a new expression of expectation value ($<\varphi_0'| \partial_x V_{23} |\varphi_0'>$ here) which can be evaluated at a lower order of R.C. With this, we can further simplify $p_y^{(1)}$, while still keeping it correct to the 1$^{st}$-order R.C., by writing

$$p_y^{(1)} = -\tau_v \frac{\hbar}{4\Delta} <\varphi_0'|\partial_x V_{23}|\varphi_0'>,$$

$$(H^{(0)} + V_3)\varphi_0' = E_0'\varphi_0',$$

where the wave equation now includes no R. C. at all. Then, applying again Ehrenfest's theorem,

$$0 = \frac{d<\varphi_0'|p_x|\varphi_0'>}{dt} = -\frac{i}{\hbar}<\varphi_0'|[p_x, H^{(0)} + V_3|\varphi_0'>,$$

one obtains

$$p_y^{(1)} \propto <\varphi_0'|\partial_x V_{23}|\varphi_0'> = 0.$$

With $p_y^{(1)} = 0$ as just shown, we rewrite (III-3),

$$<\varphi_0(x,y)|p_y\varphi_0(x,y)> \approx p_y^{(2,1)} + p_y^{(2,2)} \qquad (B-1)$$

$$\propto \tau_v k_{3x}\left(\frac{\hbar w_0}{\Delta}\right)^2,$$

expressing explicitly that the momentum matrix element and, hence, $\gamma_0$ as well are finite only at the 2$^{nd}$-order R.C.

Before we move on to calculate the momentum matrix elements remaining in (B-1), we summarize the useful trick employed above in the derivation of $p_y^{(1)}$, since it is to be utilized again in the evaluation of these matrix elements. Namely, we utilize Ehrenfest's theorem to extract out $\tau_v$, $k_{3x}$, and the order of R.C. as well, from the expectation value being evaluated, and then proceed with the perturbation theory at a lower order to calculate the new expectation value appearing after the extraction.

We calculate $p_y^{(2,1)}$ in (B-1) now. Because $p_y^{(2,1)}$ derives from $H_\tau^{(1)}$, in the 2$^{nd}$-order perturbative treatment of H' (or, rather, the part of $H^{(1)}$ only), we write

$$p_y^{(2,1)} = <\varphi_0'| p_y |\varphi_0'>,$$
$$\left(H^{(0)} + V_3 + H^{(1)}\right)\varphi_0' = E_0'\varphi_0'.$$

In addition, being linear in $k_{3x}$, $p_y^{(2,1)}$ is 1$^{st}$-order in $V_3$. Altogether, the calculation of $p_y^{(2,1)}$ would be a 3$^{rd}$-order perturbative treatment, if one calculates $p_y^{(2,1)}$ straightforwardly using the eigenstates of $H_0$ (which are harmonic oscillator wave functions) and treating both $V_3$ and $H^{(1)}$ as perturbations. We have evaluated $p_y^{(2,1)}$ in this lengthy way. On the other hand, an alternative calculation has been developed which agrees with the lengthy one but reduces the treatment to a 2$^{nd}$-order perturbative calculation, based on the trick summarized earlier. Below, we present the 2$^{nd}$ method.

We apply the trick and write

$$<\varphi_0'| p_y |\varphi_0'>$$
$$= -\frac{m^*}{i\hbar} <\varphi_0'| [y, H^{(1)}] |\varphi_0'>$$
$$= -m^* <\varphi_0'| \partial_{p_y} H^{(1)} |\varphi_0'> = p_3^{(2,1)} + p_\tau^{(2,1)} + p_0^{(2,1)},$$

(Ehrenfest's theorem: $0 = \frac{d<\varphi_0'|y|\varphi_0'>}{dt}$
$= <\varphi_0'|[y, H^{(0)} + V_3 + H^{(1)}]|\varphi_0'>.$)

$$p_3^{(2,1)} = -m^* <\varphi_0'| \partial_{p_y} H_{03}^{(1)} |\varphi_0'> = 0, \quad (B-2)$$
$$p_\tau^{(2,1)} = -m^* <\varphi_0'| \partial_{p_y} H_\tau^{(1)} |\varphi_0'> = -\tau_v \frac{\hbar}{4\Delta} <\varphi_0'| \partial_x V_{23} |\varphi_0'>,$$
$$p_2^{(2,1)} = -m^* <\varphi_0'| \partial_{p_y} H_{02}^{(1)} |\varphi_0'>.$$

We focus now on the last two matrix elements introduced in (B-2). Firstly, we reduce $p_\tau^{(2,1)}$ by writing

$$p_\tau^{(2,1)} = -\tau_v \frac{\hbar}{4\Delta} <\varphi_0'|\partial_x V_{23}|\varphi_0'>,$$
$$\left(H^{(0)} + V_3 + H_{02}^{(1)} + H_{03}^{(1)}\right)\varphi_0' = E_0'\varphi_0'.$$

Utilizing Ehrenfest's theorem, we have

$$p_\tau^{(2,1)} = -\tau_v \frac{\hbar}{4\Delta} <\varphi_0'|\partial_x V_{23}|\varphi_0'> \qquad (B-2.1)$$
$$= \tau_v \frac{\hbar}{4\Delta} <\varphi_0'|\partial_x H_{03}^{(1)}|\varphi_0'> = -\tau_v \frac{\hbar}{16} \frac{\hbar^2 w_0^2}{\Delta^2} k_{3x}.$$

We turn to the calculation of $p_2^{(2,1)}$ in (B-2). We start by reducing $p_2^{(2,1)}$,

$$p_2^{(2,1)} = -m^* <\varphi_0'|\partial_{p_y} H_{02}^{(1)}|\varphi_0'>,$$
$$\left(H^{(0)} + V_3 + H_\tau^{(1)}\right)\varphi_0' = E_0'\varphi_0'.$$

Note that $\varphi_0'$ here needs to be evaluated only to the 2$^{nd}$ order in the perturbation $V_3 + H_\tau^{(1)}$. This can be carried out, yielding

$$p_2^{(2,1)} = -\tau_v \frac{s}{64}\left[-2 - \frac{16}{3(s+1)} + \frac{50}{3(s+2)}\right]\hbar \frac{\hbar^2 w_0^2}{\Delta^2} k_{3x}. \qquad (B-2.2)$$

Collecting the results in (B-2), (B-2.1), and (B-2.2), we obtain

$$p_y^{(2,1)} = -\tau_v \frac{s}{64}\left[-2 + \frac{4}{s} - \frac{16}{3(s+1)} + \frac{50}{3(s+2)}\right]\hbar \frac{\hbar^2 w_0^2}{\Delta^2} k_{3x}. \qquad (B-3)$$

Last, we evaluate $p_y^{(2,2)}$ in (B-1). Because it derives from $H_\tau^{(2)}$ in the 1$^{st}$-order perturbative treatment of H', we write

$$p_y^{(2,2)} = <\varphi_0'|p_y|\varphi_0'>,$$
$$\left(H^{(0)} + V_3 + H_\tau^{(2)}\right)\varphi_0' = E_0'\varphi_0'.$$

Utilizing Ehrenfest's theorem, we write

$$p_y^{(2,2)} = -m^* <\varphi_0'|\partial_{p_y} H_\tau^{(2)} |\varphi_0'> = p_2^{(2,2)} + p_3^{(2,2)}, \tag{B-4}$$

$$p_2^{(2,2)} = -m^* <\varphi_{02}'|\partial_{p_y} H_{\tau 2}^{(2)} |\varphi_{02}''>,$$

$$(H^{(0)} + V_3)\varphi_{02}' = E_{02}'\varphi_{02}',$$

$$p_3^{(2,2)} = -m^* <\varphi_{03}'|\partial_{p_y} H_{\tau 3}^{(2)} |\varphi_{03}'>,$$

$$H^{(0)}\varphi_{03}' = E_{03}'\varphi_{03}'.$$

Two matrix elements, $p_2^{(2,2)}$ and $p_3^{(2,2)}$, have been introduced here. $p_2^{(2,2)}$ can be calculated with $\varphi_{02}'$ evaluated perturbatively to the 1$^{st}$ order of $V_3$, yielding

$$p_2^{(2,2)} = \tau_v \frac{[9s+1]}{64} \hbar \frac{\hbar^2 w_0^2}{\Delta^2} k_{3x}. \tag{B-4.1}$$

As for $p_3^{(2,2)}$, with $\varphi_{03}'$ being the ground state harmonic oscillator wave function, it can be calculated straightforwardly, yielding

$$p_3^{(2,2)} = -\tau_v \frac{3(3s-1)}{64} \hbar \frac{\hbar^2 w_0^2}{\Delta^2} k_{3x}. \tag{B-4.2}$$

Collecting the results in (B-4), (B-4.1), and (B-4.2), we obtain

$$p_y^{(2,2)} = \tau_v \frac{1}{16} \hbar \frac{\hbar^2 w_0^2}{\Delta^2} k_{3x}. \tag{B-5}$$

In summary, we have derived and listed in (B-1), (B-3), and (B-5) all the momentum matrix elements appearing in (III-2.2), for the evaluation of the geometric phase rate of change ($\gamma_0$).

**Appendix C**

**Coherence time of valley pair qubit in the in-plane magnetic field configuration**

We estimate the coherence time of valley pair qubits. In the absence of a normal magnetic field, the valley states are degenerate, and the qubit decoherence derives primarily from the elastic intervalley scattering K ↔ K' in each QD. Accordingly, the coherence time is determined by the following valley flip rate

$$\frac{\hbar}{\tau_{K \leftrightarrow K'}} \approx O(V_{K \leftrightarrow K'}). \tag{C-1}$$

Here, $\tau_{K \leftrightarrow K'}$ is the valley flip time, and

$$V_{K \leftrightarrow K'} = \text{intervalley coupling} = <K_D|V_{QD}|K_D'>, \tag{C-2}$$

$K_D$ and $K_D'$ being the quantized valley states in the QD. We write

$$|K_D> = \phi_K(\vec{r})e^{i\vec{K}\cdot\vec{r}}u_K, \quad |K_D> = \phi_{K'}(\vec{r})e^{i\vec{K'}\cdot\vec{r}}u_{K'}, \tag{C-3}$$

where $\Phi_K$ and $\Phi_{K'}$ are envelope functions, and $u_K$ and $u_{K'}$ are the Bloch cell-periodic functions at K and K' points, respectively. Eqns. (C-1) and (C-2) show that $\tau_{K \leftrightarrow K'}$ depends on the profile of $V_{QD}$, which provides in the intervalley scattering the large wave vector difference ($\delta k$) between $|K_D>$ and $|K_D'>$ (with $\delta k \sim O(|\mathbf{K} - \mathbf{K'}|) \sim O(A^{-1})$).

In the following, we consider the regime where $\delta k \, L \gg 1$ (L = QD size), and estimate $V_{K \leftrightarrow K'}$ in (C-2) with three profiles of $V_{QD}$. We take L (QD size) ~ 300A and potential depth $V_0 \sim 0.1 eV$, for typical applications of valley pair qubits.[5] We employ the following approximation

$$V_{K \leftrightarrow K'} = <K_D'|V_{QD}|K_D> \approx O\left(\frac{1}{L^2}\int_{r \leq L} V_{QD}(\vec{r})e^{-i\delta\vec{k}\cdot\vec{r}}d^2r\right), \tag{C-4}$$

$$\vec{\delta k} = \vec{K} - \vec{K'},$$

where the envelope functions $\Phi_K$ and $\Phi_{K'}$ in (C-4) are taken approximately to be constant for r < L and zero for r > L. Eqn. (C-4) is basically the Fourier transform of $V_{QD}$ at $\delta \mathbf{k}$, and suffices for the order-of-magnitude estimate of $V_{K \leftrightarrow K'}$.

Firstly, we consider a) an ideal square well, with barrier height $V_0$. This is the worst case scenario, since the abrupt change in the potential leads to sizable Fourier components at large wave vectors and causes frequent intervalley scattering, resulting in short $\tau_{K \leftrightarrow K'}$. A brief dimensional analysis with (C-4) in the case shows that

$$V_{K \leftrightarrow K'} \approx O\left(\frac{V_0}{(L \delta k)^2}\right) \approx O(\mu eV),$$

or

$$\tau_{K \leftrightarrow K'} \approx O(10 ns). \qquad (C-5a)$$

The coherence time is close to the time needed for VOI-based manipulation of qubits (~O(10ns)).

Next, we consider b) a realistic square well, with a transition region between well and barrier. For the convenience of analysis, we simulate $V_{QD}$ in this case with a factorizable form, such that the integral in (C-4) can be decomposed into the product of two independent one-dimensional integrals. We take $V_{QD} = V_{1/2}(x)V_{1/2}(y)$, $V_{1/2}(x) = (V_0)^{1/2}$, for -L' < x < L'; $V_{1/2}(x) = (V_0)^{1/2} (x+L)/(L- L')$, for -L < x < -L'; $V(x) = (V_0)^{1/2} (L-x)/(L- L')$, for L' < x < L; and V(x) = 0 elsewhere. $V_{1/2}(y) = - V_{1/2}(x)|_{x \to y}$. We also take O(L') ~ O(L), and the transition layer thickness $\delta L$ (= L-L') ~ 0.1L. With this factorizable potential, we obtain

$$V_{K \leftrightarrow K'} = -(\overline{V}_{1/2})^2,$$

$$\overline{V}_{1/2} = O\left(\frac{1}{L}\int_{-L}^{L} V_{1/2}(x) e^{-i\delta k \cdot x \hat{x}} dx\right).$$

Analysis of the one-dimensional integral appearing here gives

$$\overline{V}_{1/2} \approx O\left(\frac{V_0^{1/2}}{(L\delta k)(\delta L\delta k)}\right),$$

$$V_{K\leftrightarrow K'} \approx O\left(\frac{V_0}{(L\delta k)^2(\delta L\delta k)^2}\right) \approx O(1neV),$$

or

$$\tau_{K\leftrightarrow K'} \approx O(0.5\mu s). \tag{C-5b}$$

The valley flip time here is long enough for qubit manipulation.

Last, we consider c) a parabolic potential, $V_{QD} = \frac{1}{2} m^* w_0^2 r^2$, with $\hbar w_0 \sim O(V_0)$. In this case, we obtain

$$V_{K\leftrightarrow K'} \approx O\left(\frac{V_0}{(L\delta k)^4}\right) \approx O(0.01neV),$$

or

$$\tau_{K\leftrightarrow K'} \approx O(50\mu s). \tag{C-5c}$$

It provides a very long coherence time for the qubit manipulation. Comparing the results in (C-5a) – (C-5c), we see that the qubit coherence time can be increased significantly with a QD profile engineering that is able to create a smooth QD confinement potential.

# Figure Captions

**Figure 1. a)** The DQD structure of a valley pair qubit. The QDs are electrostatically defined, for example, by back gates (not shown in the figure). Gate $V_C$ is used to tune the potential barrier and also generate a linear term (in x) in the QD confinement potential. The coupling between the two QDs is characterized by the tunneling amplitude $t_{d-d}$, and may be controlled by $V_c$ or the back gates. Gates $V_L$ and $V_R$ are AC biases. **b)** A static in-plane magnetic field is applied to freeze the electron spin, leaving only the valley degree of freedom for qubit implementation. **c)** Valley singlet ($z_S$) / triplet ($z_{T0}$, $z_{T+}$, $z_{T-}$) states constitute the low energy sector of the two-electron states, with the singlet-triplet splitting being given by J.

**Figure 2.** Single qubit manipulation, with the initial qubit state, e.g., $|z_S\rangle$. One may apply the alternating sequence, $R_x(\theta_x) \to R_z(\theta_z=\pi) \to R_x(-\theta_x) \to R_z(\theta_z=\pi) \to \ldots R_z(\theta_z^{(target)}+\pi/2)$, and manipulate the initial state into a target state ($\theta_z^{(target)}$ = *target state longitude*).

**Figure 3.** The approximate selection rule of optical excitation in gapped graphene.

**Figure 4.** The quantum state transfer from a photon qubit to a valley pair qubit. Dashed circles - quantum dots; black dots – electrons; white dots – holes. Double arrows indicate resonance between states. For simplicity, in the state $\Phi_2$ plotted here, we show the approximate polarization only, for the photon emitted by the exciton, as determined by the approximate selection rule in (I-1).

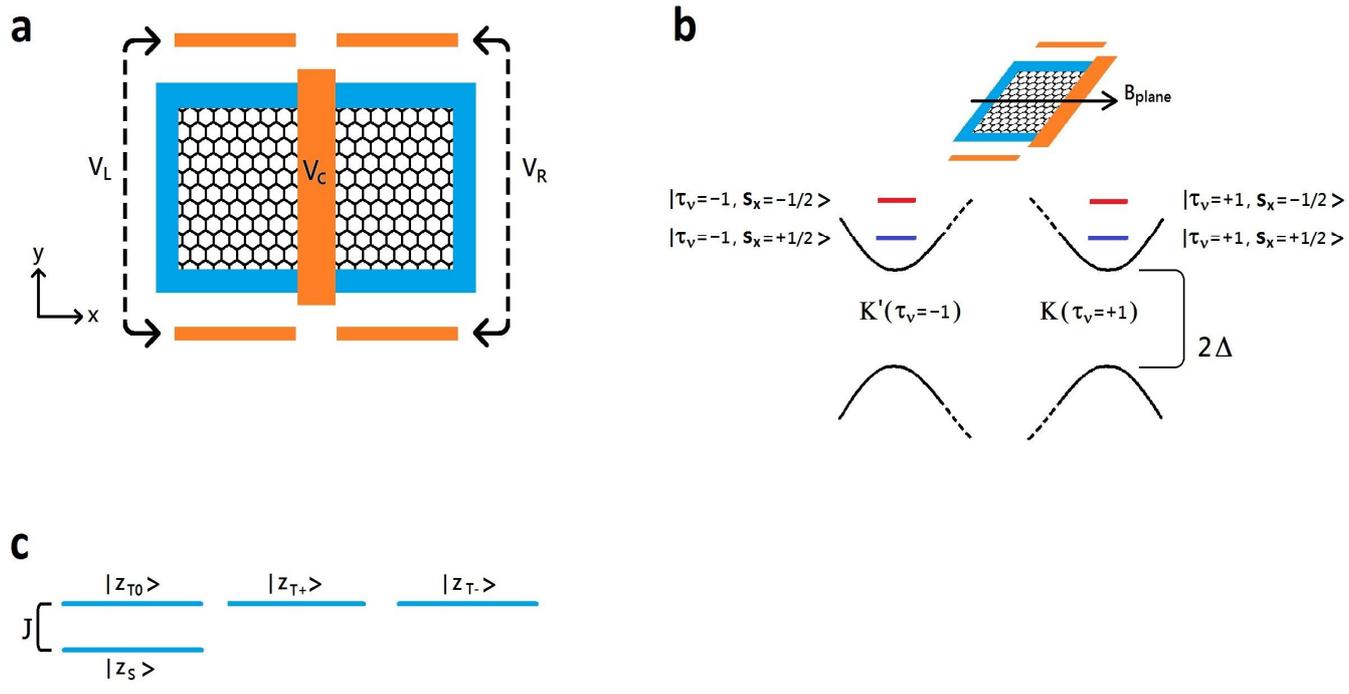

**Figure 1**

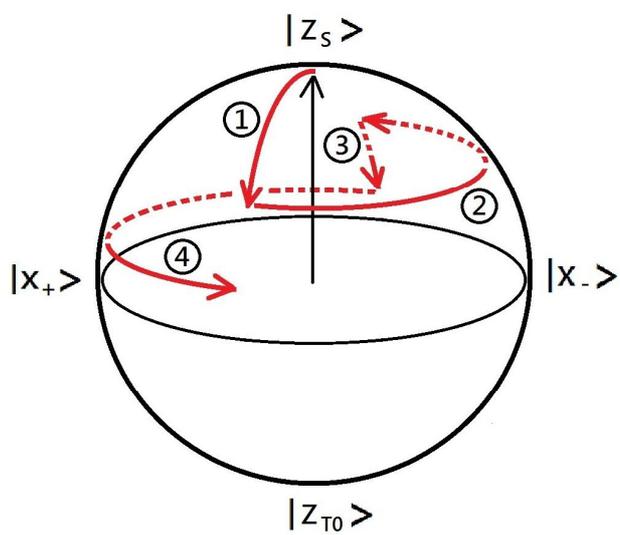

**Figure 2**

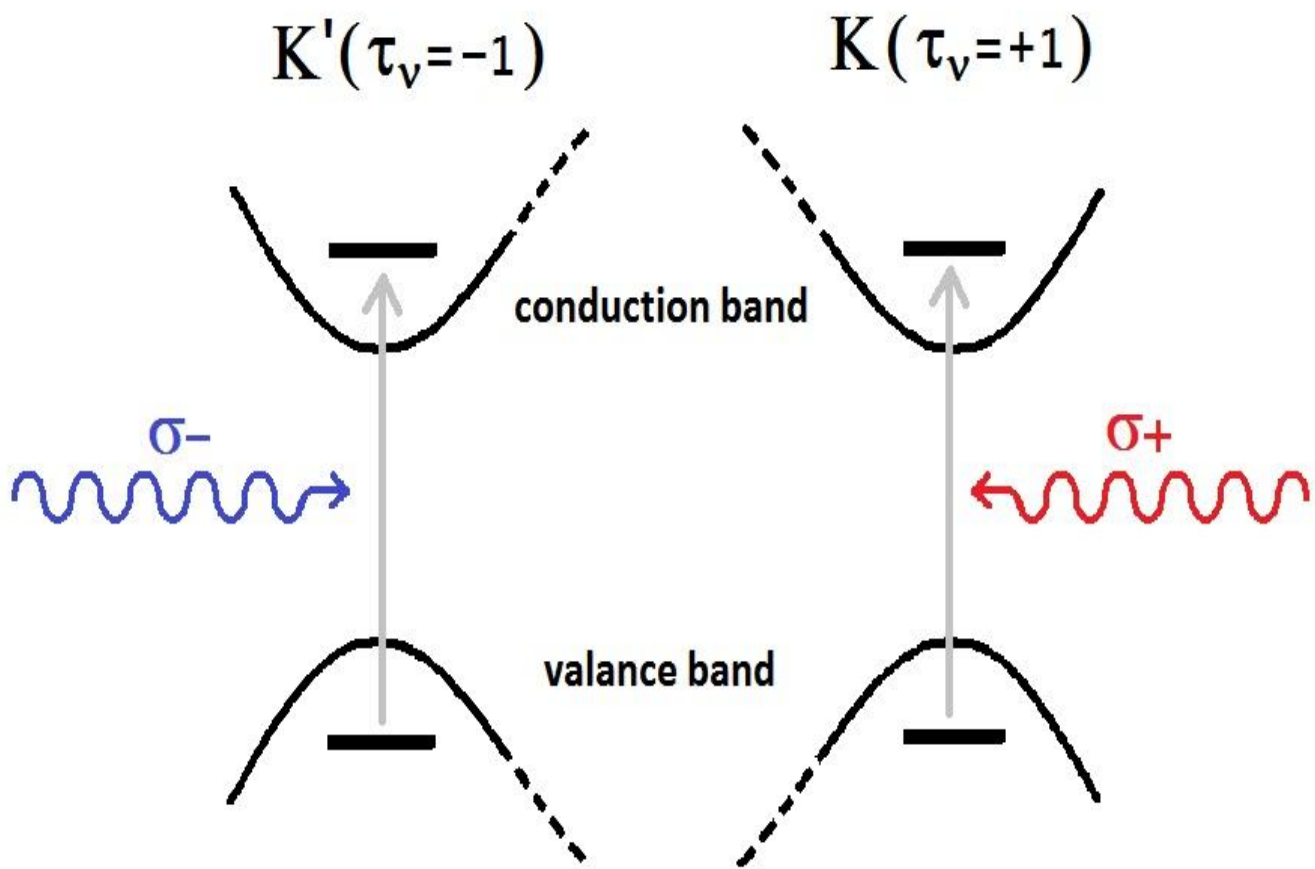

**Figure 3**

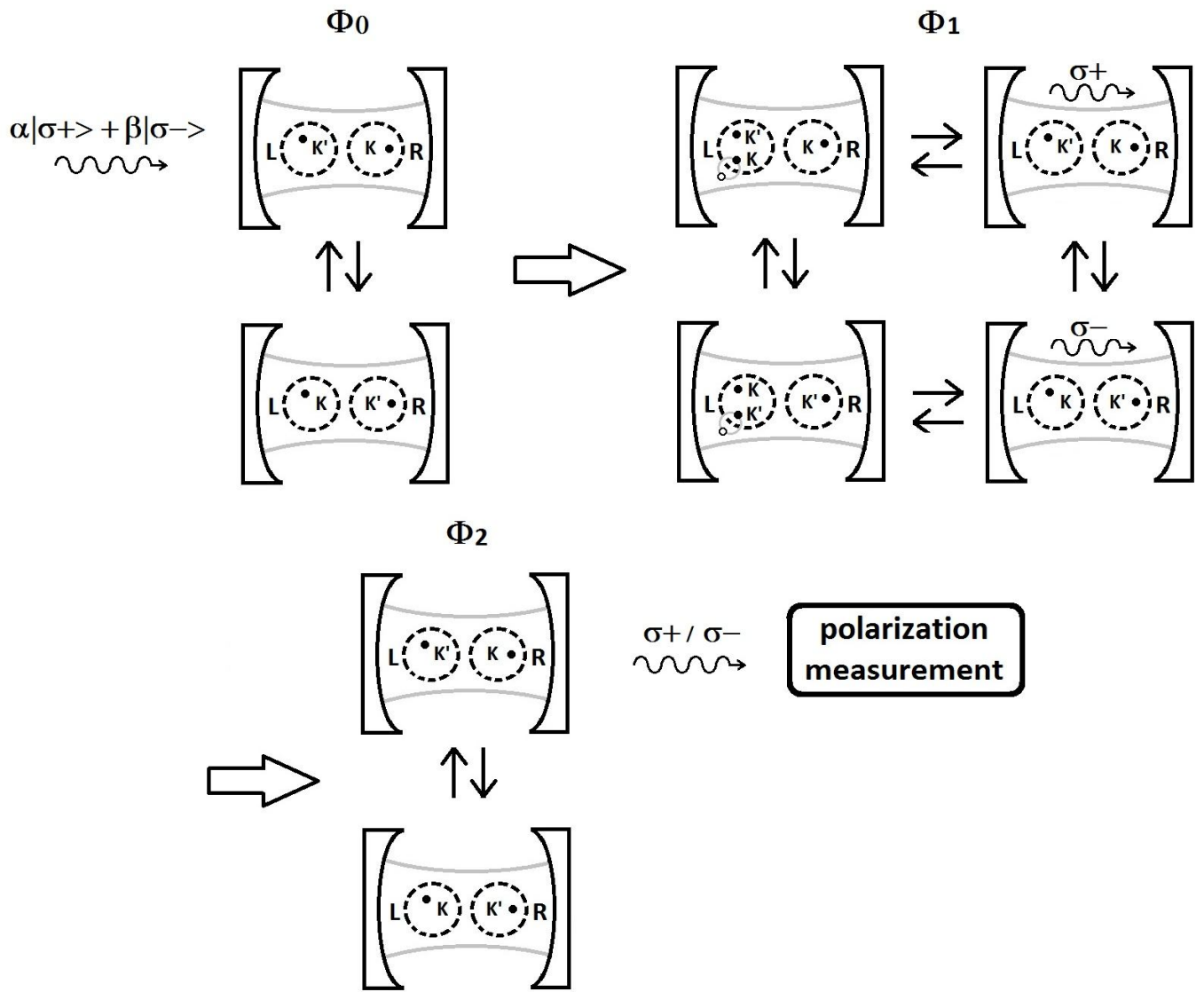

**Figure 4**